\documentclass[12pt]{iopart} 
\usepackage{iopams,isolatin1,graphicx,times}

\newcommand{\be}{\begin{equation}}
\newcommand{\ee}{\end{equation}}
\newcommand{\bea}{\begin{eqnarray}}
\newcommand{\eea}{\end{eqnarray}}

\newcommand{\kommentar}[1]{}

\begin{document}

\fontsize{12pt}{21pt}\selectfont

\title{Quantum transport on two-dimensional regular graphs}

\author{
Antonio Volta, Oliver M{\"u}lken,
and Alexander Blumen}

\address{
Theoretische Polymerphysik, Universit\"at Freiburg,
Hermann-Herder-Stra\ss{}e 3, D-79104 Freiburg, Germany}
\ead{oliver.muelken@physik.uni-freiburg.de}

\today
\begin{abstract}
We study the quantum-mechanical transport on two-dimensional graphs by means of continuous-time quantum walks and analyse the effect of different boundary conditions (BCs). For periodic BCs in both directions, i.e., for tori, the problem can be treated in a large measure analytically. Some of these results carry over to graphs which obey open boundary conditions  (OBCs), such as cylinders or rectangles. Under OBCs the long time transition probabilities (LPs) also display asymmetries for certain graphs, as a function of their particular sizes. Interestingly, these effects do not show up in the marginal distributions, obtained by summing the LPs along one direction.
\end{abstract}
\submitto{\JPA}
\maketitle

\section{Introduction}

The last decade has seen a strong upsurge in the study of quantum mechanical transport processes, especially based on extensions of
their classical counterparts \cite{Aharonov1993,farhi1998,kempe2003,Childs2004,shenvi2003}. On one hand this is due to the growing
attention being given to quantum information and to potential quantum
computers \cite{nielsen}; on the other hand, the interest is also fuelled by recent
experimental breakthroughs in the coherent energy transfer over atomic and molecular
systems (especially at extremely low temperatures) \cite{westermann2006a,varnavski2002a},
processes which, say in the framework of Frenkel excitons, are very closely
related to classical random walk models \cite{streitwieser}.\\ Previous work \cite{mb2005a,mb2005b,mvb2005a,mbb2006a} has highlighted the close relations between the classical continuous-time random walks (CTRWs) \cite{montroll1969a,blumen1984a,sokolov2002a} and
the continuous-time quantum walks (CTQWs). One difference between CTRWs and CTQWs lies the fact that
quantum-mechanically time is connected to the imaginary constant. Hence, the basic
physical aspects of the classical and the quantum picture differ, although with regards to the spatial
coordinates one finds in both cases (discrete) versions of the Laplace
operator. Hence, many of the mathematical tools used to handle Laplacian forms
in classical physics may well be used to bear fruit in the quantum mechanical
environment. Furthermore, for simple underlying geometries the quantum problem
is directly related to well-known models in polymer and in solid state
physics. For instance, walks on (one-dimensional) chains are readily treated
by a Bloch ansatz \cite{mb2005b,ziman,kittel}, when periodic boundary conditions (PBCs) are implemented.
On the other hand, open boundary conditions (OBCs) appear naturally in polymer
physics in relation to the Rouse-model \cite{Doi-Edwards}.\\In this work we focus on CTQWs over two-dimensional, \emph{finite} networks. These are topologically quite simple systems which, however, can display complex quantum mechanical features. By keeping the systems as simple as possible, we highlight the complex behaviour of the quantum mechanical transport compared to the classical one. Now, assuming either PBCs or OBCs (separately for each
direction) leads to structures topologically equivalent to
toroidal, cylindrical, and finite rectangular networks. For such networks the
eigenvalues of the system can be obtained in simple, closed form.
The average transition probabilities between distinct
sites display quite unexpected, odd behaviours in the long-time limit for some special network sizes. These findings extend the results previously
obtained for square networks \cite{mvb2005a} to rectangular ones. As an
additional feature, we analyse the average probability to be still or again at
the initial site; in this \ case one has an (analytical) lower bound, that in
many ways is quite close to the numerically established behaviour. The
advantage of the lower bound expression is that it depends only on the
eigenvalues, but not on the eigenvectors of the system. \\
The paper is structured as follows: In the next section we recall the general
properties of CTQWs on networks. In Sec.~\ref{bound} we analyse the role of the PBCs and OBCs on the transport. Section \ref{average} is devoted to the
average probability to be still or again at the initial site, both in the
classical and in the quantum case. In Sec.~\ref{limprobgen} we analyse the long time behaviour of the transport between pairs of sites, first for square networks, and then for general,
rectangular networks. 
We summarise the obtained results
in Sec.~\ref{conclusion}.
\section{Continuous time quantum walks on graphs}  \label{ctqw}
\begin{figure}[!ht]
\centerline{\includegraphics[clip=,angle=90,width=0.9\columnwidth]{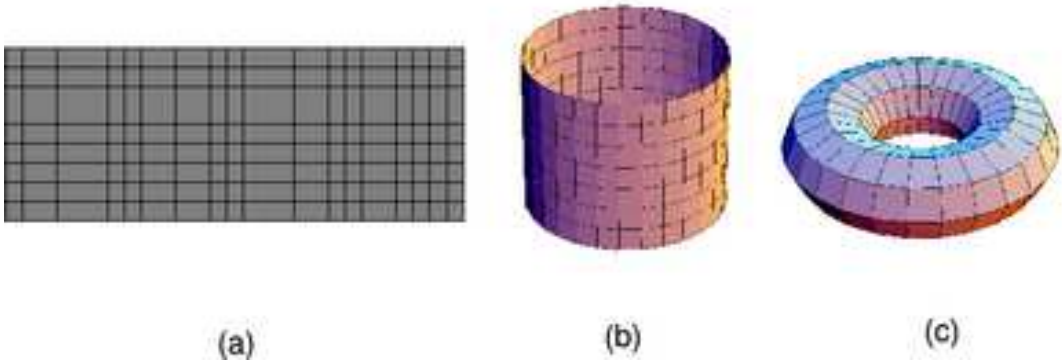}}
\caption{Two-dimensional networks with (a) finite rectangular, (b) cylindrical and (c) toroidal topology. These drawings do not exhaust the broad panoply of geometrically possible realizations of such networks and are meant solely to display clearly the connectivities.
}
\label{compact}
\end{figure}

In this work we focus on coherent quantum mechanical transport over systems whose topology can be modelled by two-dimensional graphs. Specifically, we will consider networks which consist of $M\times N=\mathcal{N}$ nodes jointed by identical bonds; such networks are topologically equivalent to finite, rectangular lattices whose length in the $x$-direction is $M$, and in the $y$-direction $N$. Then we denote the position of node $\boldsymbol{j}$ by $(j_x,j_y)$, where $j_x$ and $j_y$ are integer labels in the two directions. We will consider different situations: in each direction we either impose PBCs or OBCs. In two dimensions this leads then to three distinct topological objects, namely to a rectangle, to a cylinder and to a torus, see Fig.~\ref{compact}.
 Furthermore, we will assume that the excitation is initially localised  at site $\boldsymbol{j}$.
Classically, CTRWs are described by the master equation \cite{weiss,vankampen} :
\be 
\frac{d}{dt}p_{\boldsymbol{k,j}}(t)=\sum_{\boldsymbol{l}}T_{\boldsymbol{kl}}p_{\boldsymbol{l,j}}(t),
\label{0}
\ee 
where $p_{\boldsymbol{k,j}}(t)$ is the conditional probability to find the walker at time $t$ at node $\boldsymbol{k}$ when starting at time 0 at node $\boldsymbol{j}$. We assume an unbiased CTRW such that the transmission rates $\gamma$ of all bonds are equal. Then the transfer matrix of the walk, $\mathbf{T}=(T_{\boldsymbol{kj}})$, is related to the connectivity matrix by $\mathbf{T}=-\gamma\mathbf{A}$, where 
$\mathbf{A}=(A_{\boldsymbol{kj}})$ is a discrete form of the Laplacian operator. One has namely $A_{\boldsymbol{ij}}=-1$ if nodes $\boldsymbol{i}$ and $\boldsymbol{j}$ are connected by a bond and $A_{\boldsymbol{ij}}=0$ otherwise; furthermore $A_{\boldsymbol{ii}}=-\sum_{\boldsymbol{i}\neq \boldsymbol{j}} A_{\boldsymbol{ij}}$ or equivalently $A_{\boldsymbol{ii}}=f_{\boldsymbol{i}}$ where $f_{\boldsymbol{i}}$ is the functionality of site $\boldsymbol{i}$.\\The quantum mechanical extension of a CTRW, the CTQW, is now defined by identifying the Hamiltonian of the system with the (classical) transfer operator, $\mathbf{H}=-\mathbf{T}$ \cite{farhi1998,Childs2004,mb2005a}. 
A convenient way to implement this idea is to start from localised states, so that state $|\boldsymbol{j}\rangle\equiv |j_x\rangle \otimes |j_y\rangle\equiv|j_x,j_y\rangle$ is localised at node $\boldsymbol{j}$. The states $|\boldsymbol{j}\rangle$ span the whole accessible Hilbert space and form an orthonormal basis set, i.e., $\langle\boldsymbol{k}|\boldsymbol{j}\rangle=\delta_{\boldsymbol{kj}}$ and $\sum_{\boldsymbol{j}}|\boldsymbol{j}\rangle\langle\boldsymbol{j}|=\mathbf{1}$, with $\mathbf{1}$ being the identity operator.  
If at the initial time $t_0=0$ only state $|\boldsymbol{j}\rangle$ is populated, then the transition amplitude $\alpha_{\boldsymbol{k,j}}(t)$ 
from state $|\boldsymbol{j}\rangle$ to state $|\boldsymbol{k}\rangle$ in time $t$ obeys the Schr\"odinger Equation (SE)
\be \label{5}
i\frac{d}{dt}\alpha_{\boldsymbol{k,j}}(t)=\sum_{\boldsymbol{l}}H_{\boldsymbol{kl}}\alpha_{\boldsymbol{l,j}}(t),
\ee  
whose solution can be formally expressed as:
\be \label{4}
\alpha_{\boldsymbol{k,j}}(t)=\langle\boldsymbol{k}|\exp({-i\mathbf{H}t})|\boldsymbol{j}\rangle. 
\ee  
The corresponding transition probability during time $t$ is $\pi_{\boldsymbol{k,j}}(t)=|\alpha_{\boldsymbol{k,j}}(t)|^2$. Nevertheless, despite the formal similarity between Eqs.~(\ref{0}) and (\ref{5}), we have classically $\sum_{\boldsymbol{k}}p_{\boldsymbol{k,j}}(t)=1$ but quantum-mechanically $\sum_{\boldsymbol{k}}|\alpha_{\boldsymbol{k,j}}(t)|^2=1$.\\

~\\We hasten to note that there is no unique way of defining a CTQW. For example, for regular networks, where all nodes have the same functionality, different choices of the Hamiltonian can give rise to the same quantum dynamics \cite{Childs2004}. Nevertheless, in what follows we will stick to directly identifying the Hamiltonian with the transfer operator, since some of the networks we will consider below are not regular. For instance, the functionality of sites of a finite rectangular network ranges from 2 for a corner site to 4 for an internal site. Sites on the boundary of a cylinder have functionality 3. A torus is regular in the sense of Ref. \cite{Childs2004}, since all sites have functionality 4. \\It is now reasonable to work in an orthonormal basis ${|\boldsymbol{q}_n\rangle}$, $n\in[1,\mathcal{N}]$ which diagonalises $\mathbf{A}$, and hence also $\mathbf{H}=-\mathbf{T}=\gamma\mathbf{A}$. For this we need to know all the eigenvalues and the eigenvectors of $\mathbf{A}$. 
As is well-known, the matrix $\mathbf{A}$ is non-negative definite; moreover, for a connected underlying structure it has only one vanishing eigenvalue, the other eigenvalues being strictly positive \cite{biswas2000}. Thus $\mathbf{A}=\mathbf{Q\Lambda Q^{-1}}$,
where $\mathbf{Q}$ is the matrix built up by the $|\boldsymbol{q}_n\rangle$, $\mathbf{Q^{-1}}$ is its inverse, and $\mathbf{\Lambda}$ 
is the diagonal matrix having as diagonal elements the eigenvalues $\lambda_{n}$ of $\mathbf{A}$.
\\ Let us now turn our attention to the probability to be still or again at the starting point of the motion $p_{\boldsymbol{k,k}}(t)$. We can write the formal solution of Eq.~(\ref{0}) in Dirac notation as
\be \label{classica}
p_{\boldsymbol{k,j}}(t)=\langle\boldsymbol{k}|\exp({\mathbf{T}t})|\boldsymbol{j}\rangle, 
\ee 
where classically the vector $|\boldsymbol{j}\rangle$ has a 1 at the position corresponding to the pair $(j_x, j_y)$ and zeros otherwise. Then, $p_{\boldsymbol{k,k}}(t)$ reads
\be \label{10}
p_{\boldsymbol{k,k}}(t)=\langle\boldsymbol{k}|\exp({\mathbf{T}t})|\boldsymbol{k}\rangle=\sum_{n}\langle \boldsymbol{k}|\boldsymbol{q}_n\rangle \exp({-\gamma t \lambda_{n}})\langle \boldsymbol{q}_n|\boldsymbol{k}\rangle.
\ee 
As also shown in \cite{alexander1981,blumen2005a,blumen2005b}, the average of the classical probability $p_{\boldsymbol{k,k}}(t)$ over all sites of the graph is
\begin{displaymath}
\bar{p}(t)=\frac{1}{\mathcal{N}}\sum_{\boldsymbol{k}}p_{\boldsymbol{k,k}}(t)
\end{displaymath}
\be \label{11}
=\frac{1}{\mathcal{N}}\sum_{n}\sum_{\boldsymbol{k}}\langle\boldsymbol{q}_n|\boldsymbol{k}\rangle\langle\boldsymbol{k}|\boldsymbol{q}_n\rangle \exp({-\gamma t \lambda_{n}})=\frac{1}{\mathcal{N}}\sum_{n}\exp({-\gamma\lambda_{n}t}).
\ee 
 The classical $\bar{p}(t)$ decays monotonically from $\bar{p}(t)=1$ to a final, asymptotic plateau, $\lim_{t\to \infty}\bar{p}(t)=1/\mathcal{N}$. This behaviour is typical for a diffusive process leading to energy equipartition.
From Eq.~(\ref{11}) it is evident that $\bar{p}(t)$ depends \textit{only} on the eigenvalues and not on the eigenvectors of $\mathbf{T}$.\\
For CTQWs, the average probability $\bar{\pi}(t)$ to be still or again at the initial state at time $t$ reads,
\begin{displaymath}
\bar{\pi}(t)=\frac{1}{\mathcal{N}}\sum_{\boldsymbol{k}}\pi_{\boldsymbol{k,k}}(t)=\frac{1}{\mathcal{N}}\sum_{\boldsymbol{k}}\left|\alpha_{\boldsymbol{k,k}}(t)\right|^2
\end{displaymath}
\be \label{13}
=\frac{1}{\mathcal{N}}\sum_{n,m}\exp[{-i\gamma\left (\lambda_{n}-\lambda_{m}\right )t}]\sum_{\boldsymbol{k}}\left|\langle \boldsymbol{k}\right|\boldsymbol{q}_n\rangle\left|^2\right|\langle \boldsymbol{k}\left|\boldsymbol{q}_m\rangle\right|^2,
\ee
where we used that $\alpha_{\boldsymbol{k,k}}(t)=\sum_n\langle\boldsymbol{k}|\boldsymbol{q}_n\rangle\langle\boldsymbol{q}_n|\boldsymbol{k}\rangle\exp({-i\gamma \lambda_nt})$.
Eq.~(\ref{13}) depends explicitly on the eigenvectors, which renders its determination cumbersome. As shown in \cite{mbb2006a}, $\bar{\pi}(t)$ admits a lower bound, as follows using the Cauchy-Schwarz inequality:
\begin{displaymath}
\bar{\pi}(t)=\frac{1}{\mathcal{N}}\sum_{\boldsymbol{k}}\pi_{\boldsymbol{k,k}}(t)=\frac{1}{\mathcal{N}}\sum_{\boldsymbol{k}}^{\mathcal{N}}\left|\alpha_{\boldsymbol{k,k}}(t)\right|^2
\end{displaymath}
\be \label{15}
=\sum_{\boldsymbol{k}}\left|\frac{1}{\sqrt{\mathcal{N}}}\alpha_{\boldsymbol{k,k}}(t)\right|^2 \sum_{\boldsymbol{k}}\left({\frac{1}{\sqrt{\mathcal{N}}}}\right)^2\geq \left|\frac{1}{\mathcal{N}}\sum_{\boldsymbol{k}}\alpha_{\boldsymbol{k,k}}(t)\right|^2\equiv\mu(t).
\ee
Furthermore, in analogy to Eq.~(\ref{11}) we have
\begin{displaymath}
\sum_{\boldsymbol{k}}\alpha_{\boldsymbol{k,k}}(t)=\sum_{\boldsymbol{k}}\sum_{n}\exp({-i\gamma\lambda_{n}t})\langle\boldsymbol{q}_n|\boldsymbol{k}\rangle\langle\boldsymbol{k}|\boldsymbol{q}_n\rangle
\end{displaymath}
\be \label{15bis}
=\sum_{n} \exp({-i\gamma\lambda_{n}t})\langle\boldsymbol{q}_n|\boldsymbol{q}_n\rangle=\sum_{n}\exp({-i\gamma\lambda_{n}t}).
\ee
Thus $\mu(t)$ equals
\begin{displaymath}
\mu(t)=\frac{1}{\mathcal{N}^2}\left|\sum_{\boldsymbol{k}}\alpha_{\boldsymbol{k,k}}(t)\right|^2=\frac{1}{\mathcal{N}^2}\left|\sum_{n}\exp({-i\gamma\lambda_{n}t})\right|^2
\end{displaymath}
\be \label{16}
=\frac{1}{\mathcal{N}^2}\sum_{n,m}\exp[{-i\gamma(\lambda_{n}-\lambda_{m})t}].
\ee
The last expression shows that also $\mu(t)$ depends only on the eigenvalues. As we will discuss in Sec.~{\ref{average}}, $\mu(t)$ provides important information about the network, despite of being only a lower bound .

\section{Role of different boundary conditions} \label{bound}
We now turn to two-dimensional networks built from $\mathcal{N}=M\times N$ nodes, which, by implementing different boundary conditions (OBCs or PBCs) stay rectangular or turn into cylinders or tori, see Fig.\ref{compact}. These symmetries are based on the form of the matrix $\mathbf{A}$ and are only topological in nature; in fact, it is not compulsory for these structures to obey any kind of translation symmetry, see for instance the situation in polymer physics \cite{gurtov2002} or in quantum chemistry \cite{blumen1977}, where the $\mathbf{A}$-matrices appear in H\"uckel molecular orbital calculations \cite{streitwieser}.
For the sake of simplicity and without any loss of generality we set now the transmission rate $\gamma=1$.\\ To fix the ideas let us first consider the one-dimensional case, namely a finite chain consisting of $M$ nodes, arranged in the $x$-direction. Under PBCs the corresponding Hamiltonian, $\mathbf{H}_x^p$, reads 
\be \label{equationchain1}
\mathbf{H}_x^p|j_x\rangle=2|j_x\rangle - |j_x-1\rangle - |j_x+1\rangle,
\ee
where this relation is to be understood modulo $M$, in particular $|M\rangle\equiv |0\rangle$ and $|M+1\rangle\equiv|1\rangle$. The corresponding eigenfunctions are $|\Psi_{\theta_x}\rangle=(1/\sqrt{M})\sum_{j_x=1}^M\exp(-i\theta_xj_x)|j_x\rangle$ and correspond to the eigenvalues $\lambda_{\theta_x}=2-2\cos{\theta_x}$, where $\theta_x=2\pi m/M$, with $m=0,1,\dots,M-1$.\\ Under OBCs the corresponding hamiltonian $\mathbf{H}_x^o$ reads: 
\be
\mathbf{H}_x^o|1\rangle=|1\rangle - |2\rangle \hspace{0.5cm},\hspace{0.5cm} \mathbf{H}_x^o|M\rangle=|M\rangle - |M-1\rangle 
\ee 
and 
\be
\mathbf{H}_x^o|j_x\rangle\equiv\mathbf{H}_x^p|j_x\rangle
\ee
for $|j_x\rangle$ with $j_x\in \lbrace 2,3,\dots, M-1\rbrace$. To determine the eigenvalues here one can proceed by employing well-known methods from polymer physics for the treatment of finite discrete chains \cite{Doi-Edwards,gurtov2005a}. It turns out that $\mathbf{H}_x^o$ also leads to eigenvalues of the form $\lambda_{\theta_x}=2-2\cos{\theta_x}$, but that now $\theta_x$ equals $\theta_x=\pi m/M$, with $m=0,1,\dots,M-1$ \cite{gurtov2005a}.\\ These results allow us a straightforward analysis of the three network types given in Fig.~\ref{compact} . Let us hence turn to considering the two-dimensional cases.

\subsection{The torus} \label{torusnet}
We start by considering a finite two-dimensional regular network with PBCs in both directions, i.e., a torus, see Fig.~\ref{compact}(c). 
In this case the Hamiltonian $\mathbf{H}^t$ reads
\begin{displaymath} 
\mathbf{H}^t|j_x,j_y\rangle=2|j_x,j_y\rangle-|j_x-1,j_y\rangle-|j_x+1,j_y\rangle
\end{displaymath}
\be \label{17}
\qquad \qquad +2|j_x,j_y\rangle-|j_x,j_y-1\rangle-|j_x,j_y+1\rangle,
\ee
where we take the labels $j_x$ and $j_y$ modulo $M$ and $N$, respectively. Hence 
\begin{displaymath}
\mathbf{H}^t|j_x,j_y\rangle\equiv\mathbf{H}^t(|j_x\rangle\otimes|j_y\rangle)
\end{displaymath}
\be \label{18}
\qquad \qquad=(\mathbf{H}_x^p|j_x\rangle)\otimes|j_y\rangle + |j_x\rangle\otimes(\mathbf{H}_y^p|j_y\rangle)\equiv (\mathbf{H}_x^p+\mathbf{H}_y^p)|j_x,j_y\rangle,
\ee
as can be seen by directly inserting Eq.~(\ref{equationchain1}) into Eq.~(\ref{18}), by which Eq.~(\ref{17}) is recovered. The result corresponds, of course, to a well-known result from solid-state-physics \cite{ziman,kittel}. Now, from $\mathbf{H}^t\equiv \mathbf{H}_x^p+\mathbf{H}_y^p$ it follows that the eigenvalues of $\mathbf{H}^t$ are given by 
\be \label{20}
\lambda_{\boldsymbol{\theta}}=\lambda_{\theta_x}+\lambda_{\theta_y}=4-2\cos\theta_x -2\cos{\theta_y}
\ee
with $\theta_x=2m\pi/M$ with $m=0,1,\dots, M-1$ and $\theta_y=2n\pi/N$ with $n=0,1,\dots, N-1$, and that the eigenfunctions are 
\be \label{19}
|\psi_{\boldsymbol{\theta}}\rangle\equiv |\psi_{\theta_x}\rangle\otimes|\psi_{\theta_y}\rangle=\frac{1}{\sqrt{\mathcal{N}}}\sum_{j_x,j_y=1}^{M,N}\exp[{-i(\boldsymbol{\theta}\cdot\boldsymbol{j})}]|\boldsymbol{j}\rangle,
\ee
where $\boldsymbol{\theta}\cdot\boldsymbol{j}=\theta_xj_x + \theta_yj_y$. By inverting Eq.~(\ref{19}), the state $|\boldsymbol{j}\rangle$ can be written as a linear combination of Bloch states $|\psi_{\boldsymbol{\theta}}\rangle$. The projection of the eigenstates $|\psi_{\boldsymbol{\theta}}\rangle$ on $|\boldsymbol{j}\rangle$, i.e., $\psi_{\boldsymbol{\theta}}(\boldsymbol{j})\equiv \langle \boldsymbol{j}|\psi_{\boldsymbol{\theta}}\rangle= \exp[{-i(\boldsymbol{\theta\cdot j})}]/\sqrt{\mathcal{N}}$, leads to $\psi_{\boldsymbol{\theta}}(j_x+1,j_y+1)=\exp[{-i(\theta_x+\theta_y)}]\psi_{\boldsymbol{\theta}}(j_x,j_y)$, which corresponds to Bloch's theorem. The Bloch eigenstates of the system are orthonormal, i.e., $\langle\psi_{\boldsymbol{\theta}}|\psi_{\boldsymbol{\theta}'}\rangle=\delta_{\boldsymbol{\theta},\boldsymbol{\theta}'}$, and they represent a complete set, i.e., $\sum_{\boldsymbol{\theta}}|\psi_{\boldsymbol{\theta}}\rangle\langle\psi_{\boldsymbol{\theta}}|=\mathbf{1}$ holds.\\ 
 The quantum mechanical transition amplitude from state $|\boldsymbol{j}\rangle$ to state $|\boldsymbol{k}\rangle$ in time $t$ now reads
\bea
\alpha_{\boldsymbol{k,j}}(t) 
&=& 
\frac{1}{\mathcal{N}} \sum_{\boldsymbol{\theta,\theta'}}
\langle\psi_{\boldsymbol\theta'} | \exp[{-i\boldsymbol\theta' \cdot
\boldsymbol k}] \exp({-i{\bf
H}t}) \exp[{i\boldsymbol \theta \cdot \boldsymbol j}]|\psi_{\boldsymbol \theta}\rangle 
\nonumber \\
&=& 
\frac{1}{\mathcal{N}} \sum_{\boldsymbol{\theta}} \exp({-i \lambda_{\boldsymbol\theta}t}) \exp[{-i
\boldsymbol \theta \cdot (\boldsymbol k - \boldsymbol j)}].
\label{transamplbloch}
\eea 
For networks of infinite size, i.e., when $M,N\to\infty$, the sums of Eq.~(\ref{transamplbloch}) turn into integrals. This leads to
\bea
\lim_{M,N\to\infty} \alpha_{\boldsymbol{k,j}}(t) 
&=& 
\frac{\exp({-i4t})}{4\pi^2}
\int\limits_{-\pi}^{\pi} d\theta_x \ \exp[{-i\theta_x(k_x-j_x)}]
\exp({i2t\cos\theta_x}) 
\nonumber \\
&& \times 
\int\limits_{-\pi}^{\pi} d\theta_y \ \exp[{-i\theta_y(k_y-j_y)}]
\exp({i2t\cos\theta_y})
\\
&=& 
i^{k_x-j_x} i^{k_y-j_y} \exp({-i4t}) J_{k_x-j_x}(2t) J_{k_y-j_y}(2t) \nonumber ,
\eea
where $J_{l}(2t)$ is the Bessel function of the first kind \cite{ito}. In agreement with previous calculations \cite{mb2005a,mb2005b,mvb2005a} the transition probability from $\boldsymbol{j}$ to $\boldsymbol{k}$ reads then
\be
\lim_{M,N\to\infty} \pi_{\boldsymbol{k,j}}(t) = [J_{k_x-j_x}(2t) J_{k_y-j_y}(2t)]^2.
\label{prob_bessel}
\ee
Remarkably, the lower bound $\mu(t)$ for the quantum mechanical probability to be still or again at the initial node [see Eq.~(\ref{16})] becomes exact for PBCs \cite{bbm2006a}. Namely, inserting Eq.~(\ref{transamplbloch}) for $\boldsymbol{k}=\boldsymbol{j}$ into Eq.~(\ref{13}) results in 
\bea
\bar{\pi}(t)
&=&
\frac{1}{\mathcal{N}^3}\sum_k\sum_{\boldsymbol{\theta,\theta'}}\exp({-i\lambda_{\boldsymbol{\theta}}t})\exp({i\lambda_{\boldsymbol{\theta'}}t})
\nonumber\\
&=&
\frac{1}{\mathcal{N}^2}\sum_{\boldsymbol{\theta,\theta'}}\exp[{-i(\lambda_{\boldsymbol{\theta}}-\lambda_{\boldsymbol{\theta'}})t}]=\mu(t).
\label {stilltorus}
\eea
\subsection{The cylinder and the rectangle}
\begin{figure}[!ht]
\centerline{\includegraphics[clip=,angle=270,width=0.9\columnwidth]{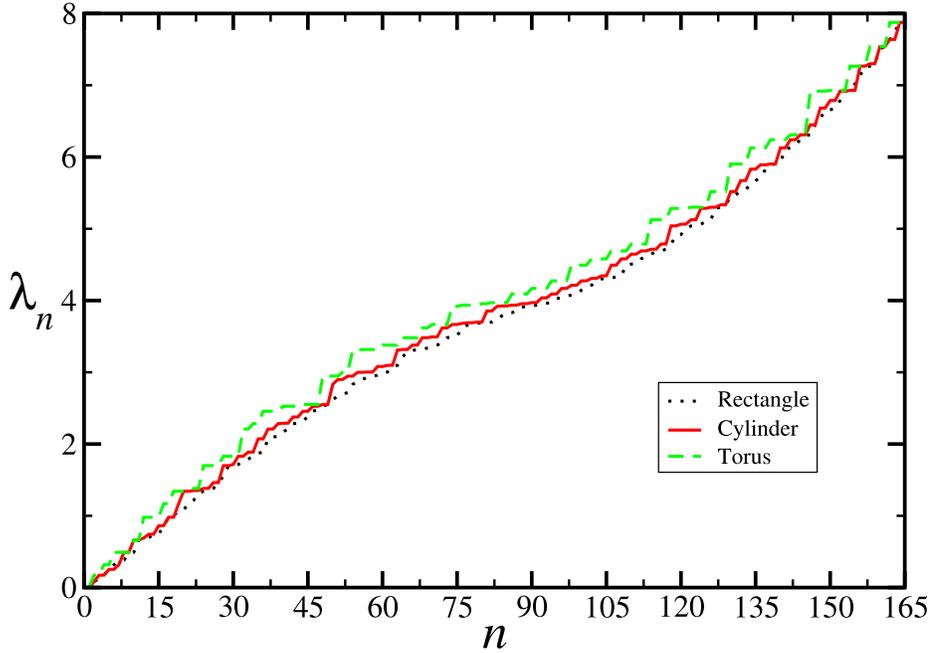}}
\caption{Eigenvalues $\lambda_n$, arranged in ascending order for the $15\times11$ rectangle, cylinder and torus, respectively.
}
\label{spectra11x15}
\end{figure}

We continue by considering now rectangular networks, on which we may or may not apply PBCs. In both cases for interior points Eq.~(\ref{17}) holds. Let us start from the cylinder, for which we impose PBCs in the $x$- and OBCs in the $y$-direction. Eq.~(\ref{17}) then also holds for all $j_x$ modulo $M$, where we identify $|M,j_y\rangle\equiv |0,j_y\rangle$ and $|1,j_y\rangle\equiv |M+1,j_y\rangle$. For sites on the upper and lower row of the cylinder one has:
\be \label{cilindro1}
\mathbf{H}^c|j_x, 1\rangle=3|j_x, 1\rangle - |j_x-1,1\rangle - |j_x+1,1\rangle - |j_x,2\rangle
\ee
and 
\be \label{cilindro2}
\mathbf{H}^c|j_x, N\rangle=3|j_x, N\rangle - |j_x-1,N\rangle - |j_x+1,N\rangle - |j_x,N-1\rangle.
\ee
It is now a simple matter to show, in analogy to Eq.~(\ref{18}) that for all $|j_x,j_y\rangle$ one has 
\be \label{cilindro3}
\mathbf{H}^c|j_x, j_y\rangle=(\mathbf{H}_x^p|j_x\rangle)\otimes|j_y\rangle + |j_x\rangle \otimes(\mathbf{H}_y^o|j_y\rangle)=(\mathbf{H}_x^p+\mathbf{H}_y^o)|j_x,j_y\rangle.
\ee 
Hence, $\mathbf{H}^c=\mathbf{H}_x^p+\mathbf{H}_y^o$ and the problem separates. Now the relation $\lambda_{\boldsymbol{\theta}}^c=\lambda_{\theta_x}^p+\lambda_{\theta_y}^o=4 - 2\cos{\theta_x} - 2\cos{\theta_y}$ holds, where $\theta_x=2\pi m/M$ with $m=0,1,\dots,M-1$ and $\theta_y=\pi n/N$ with $n=0,1, \dots,N-1$.\\ Analogously, for a rectangle, we have OBCs both in the $x$- and in the $y$-direction. Let us denote the corresponding hamiltonian by $\mathbf{H}^r$. For $\mathbf{H}^r$ the right-hand side of Eq.~(\ref{17}) holds for the internal nodes and the right-hand sides of Eqs.~(\ref{cilindro1}) and (\ref{cilindro2}) hold (apart from the corners) for the upper and lower rows. Furthermore, similar expressions hold for the nodes on the left side and on the right side of the rectangle (again, excluding the corners). This means that for all nodes considered, we have 
\be\label{cilindro4}
\mathbf{H}^r|j_x, j_y\rangle=(\mathbf{H}_x^o+\mathbf{H}_y^o)|j_x,j_y\rangle.
\ee 
Now it remains to be shown that Eq.~(\ref{cilindro4}) holds also for the corners. Exemplarily, we consider the corner, $|1,1\rangle$, for which one has
\begin{displaymath}
\mathbf{H}^r|1,1\rangle=|1,1\rangle- |2,1\rangle + |1,1\rangle - |1,2\rangle
\end{displaymath}
\be\label{cilindro6}
=(\mathbf{H}_x^o|1\rangle)\otimes |1\rangle + |1\rangle\otimes(\mathbf{H}_y^o|1\rangle)=(\mathbf{H}_x^o + \mathbf{H}_x^o)|1,1\rangle.
\ee  
This completes the proof of Eq.~(\ref{cilindro4}) for all $|j_x, j_y\rangle$, which means that $\mathbf{H}^r=\mathbf{H}_x^o+\mathbf{H}_y^o$ and thus the problem separates. Hence, the eigenvalues of $\mathbf{H}^r$ are given by $\lambda_{\boldsymbol{\theta}}^r=\lambda_{\theta_x}^o+\lambda_{\theta_y}^o=4-2\cos{\theta_x}-2\cos{\theta_y}$, where $\theta_x=\pi m/M$ with $m=0,1,\dots,M-1$ and $\theta_y=\pi n/N$ with $n=0,1, \dots,N-1$.\\
Figure \ref{spectra11x15} displays the eigenvalues for a rectangle, for a cylinder and for a torus containing each $15\times11$ sites. Now, in each case the eigenvalues lie with in interval [0,8[, see e.g. Eq.~(\ref{20}). Figure \ref{spectra11x15}, moreover, shows nicely that the $n$-th eigenvalue of the rectangle is always smaller than or equal to the $n$-th eigenvalue of the cylinder, which in turn is smaller than or equal to the $n$-th eigenvalue of the torus.

\section{Probability of being at the original site} \label{average}
\begin{figure}[!ht]
\centerline{\includegraphics[clip=,angle=270,width=0.7\columnwidth]{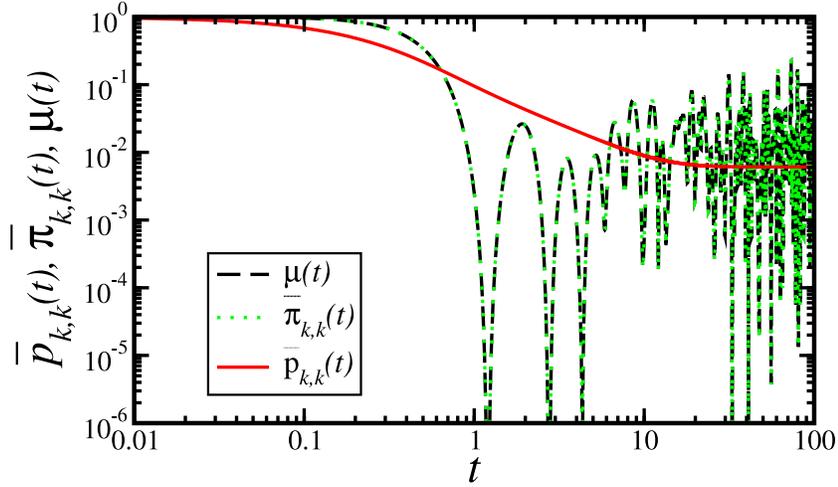}}
\caption{Classical, $\bar{p}_{\boldsymbol{k,k}}(t)$, and quantum mechanical, $\bar{\pi}_{\boldsymbol{k,k}}(t)$, probabilities to be still or again at the initial site and the lower bound $\mu(t)$ defined by Eq.~(\ref{16}) for the $15\times11$ torus.
}
\label{still11x15tor}
\end{figure}

Classically, the average probability to be still or again at the initial site is given by $\bar{p}(t)$ [Eq.~(\ref{11})]. For our structures, we can get $\bar{p}(t)$ without numerically diagonalising $\mathbf{A}$, because $\bar{p}(t)$ depends only on the eigenvalues, which are exactly known from Eq.~(\ref{20}) (with different values for $\theta_x$ and $\theta_y$, depending on the BCs of the structure). Quantum-mechanically the corresponding expression is $\bar{\pi}(t)$ [Eq.~(\ref{15})], to which $\mu(t)$ [Eq.~(\ref{16})] was shown to be a lower bound. We recall that $\bar{\pi}(t)$ depends also on the eigenvectors, whereas $\mu(t)$ does not. In the following figures we consider as an example the $15\times 11$ network with the three different BCs.
We start from the torus, i.e., from PBCs in both directions. Figure \ref{still11x15tor}, in which we plot $\bar{p}(t)$, $\bar{\pi}(t)$ and $\mu(t)$, confirms the fact that for PBCs $\bar{\pi}(t)$ and $\mu(t)$ coincide [see Eq.~(\ref{stilltorus})]. In the intermediate range  (in Fig.~\ref{still11x15tor} from $t=0.5$ to $t=5$) the classical probability $\bar{p}(t)$ is algebraic, i.e., here we have $\bar{p}(t)\sim t^{-1}$.
 
\begin{figure}[!ht]
\centerline{\includegraphics[clip=,angle=270,width=0.7\columnwidth]{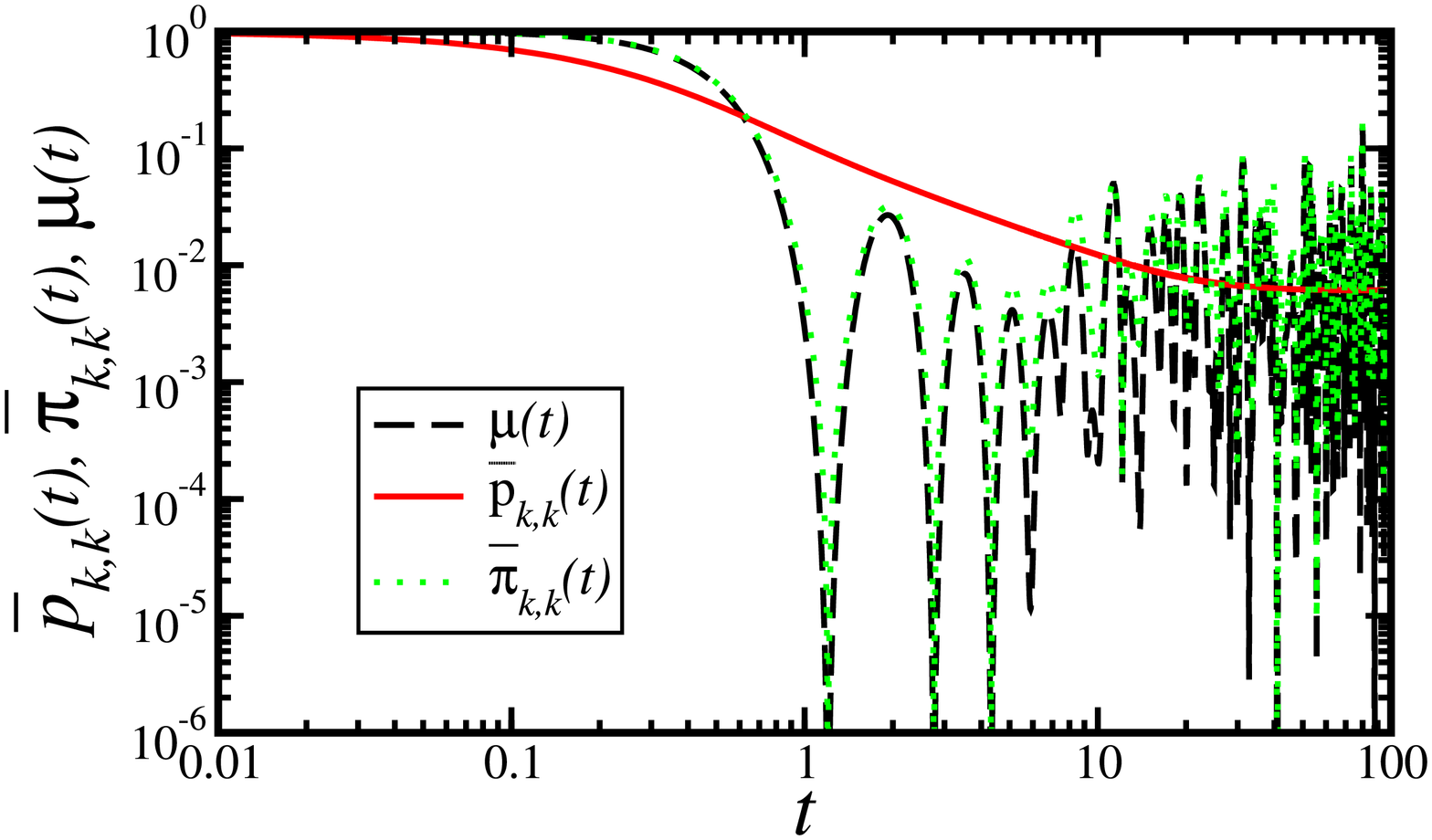}}
\caption{Classical, $\bar{p}_{\boldsymbol{k,k}}(t)$, and quantum mechanical, $\bar{\pi}_{\boldsymbol{k,k}}(t)$, probabilities to be still or again at the initial site and the lower bound $\mu(t)$ defined by Eq.~(\ref{16}) for the $15\times11$ cylinder.
}
\label{still11x15cil}
\end{figure}

\begin{figure}[!ht]
\centerline{\includegraphics[clip=,angle=270,width=0.7\columnwidth]{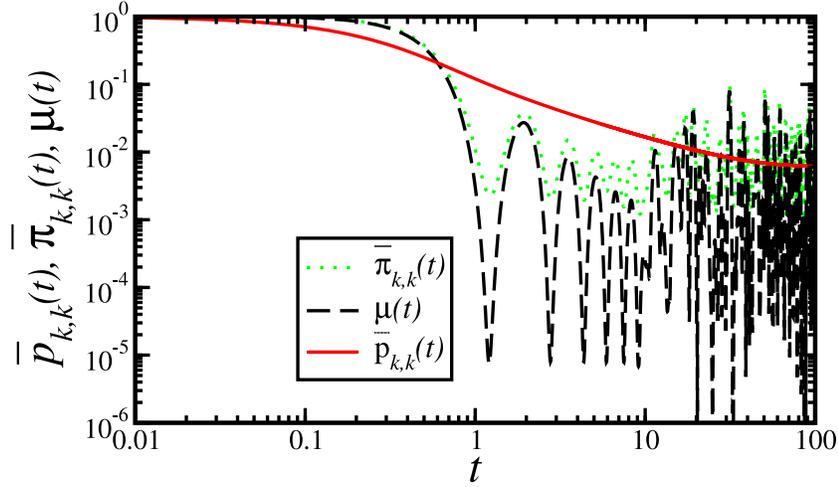}}
\caption{Classical, $\bar{p}_{\boldsymbol{k,k}}(t)$, and quantum mechanical, $\bar{\pi}_{\boldsymbol{k,k}}(t)$, probabilities to be still or again at the initial site and the lower bound $\mu(t)$ defined by Eq.~(\ref{16}) for the $15\times11$ rectangle. 
}
\label{still11x15}
\end{figure}

~\\
Figures \ref{still11x15cil} and \ref{still11x15} show for the cylinder and for the rectangle, respectively, the probability to be still or again at the initial site in the classical and in the quantum case. Furthermore, we also show the quantum mechanical lower bound $\mu(t)$, which now differs from $\bar{\pi}(t)$. However, depending on the BCs, this difference is very small and the maxima and the minima of $\mu(t)$ and $\bar{\pi}(t)$ occur basically at the same times. By comparing Fig.~\ref{still11x15tor} to Fig.~\ref{still11x15cil} and to Fig.~\ref{still11x15} one can conclude that $\bar{p}(t)$ reaches the asymptotical behaviour given by $\bar{p}(t)=1/\mathcal{N}$ earlier for the torus than for the cylinder and for the rectangle. We explain this finding by the fact that some sites on the torus are more connected than the corresponding sites on the cylinder or on the rectangle.

\section{Limiting probability distributions}\label{limprobgen}

The unitary time evolution prevents the quantum mechanical transition probability from having a definite limit when $t\to\infty$. For comparison to the classical long time probability, the long time average of $\pi_{\boldsymbol{k,j}}$, i.e., 
\be \label{9}
\chi_{\boldsymbol{k,j}}\equiv\lim_{T\to\infty}\frac{1}{T}\int_0^T dt\hspace{0.2cm}\pi_{\boldsymbol{k,j}}(t), 
\ee
can be used, see for instance \cite{aharonov2001}. 
In order to avoid the numerical integration, which requires long computing times already for moderately large network sizes, we rewrite the LP as follows \cite{mvb2005a}
\bea
\chi_{\boldsymbol{k,j}} &=
\lim_{T\to\infty}\frac{1}{T} \int\limits_0^T dt \ \left|
\sum_{n}\langle \boldsymbol k | \exp({-i{\bf H}t}) |
\boldsymbol q_n \rangle \langle \boldsymbol q_n |
\boldsymbol j \rangle \right|^2 
\nonumber \\
&=
\sum_{n,m} \langle \boldsymbol k |
\boldsymbol q_n \rangle \langle \boldsymbol q_n |
\boldsymbol j \rangle \langle \boldsymbol j | \boldsymbol q_m
\rangle \langle \boldsymbol q_m | \boldsymbol k \rangle
\left( \lim_{T\to\infty}\frac{1}{T} \int\limits_0^T dt \
\exp[{-i(\lambda_n - \lambda_m) t}] \right)
\nonumber
\\
&=
\sum_{n,m} 
\delta_{\lambda_n,\lambda_m}
\langle \boldsymbol k |
\boldsymbol q_n \rangle \langle \boldsymbol q_n |
\boldsymbol j \rangle \langle \boldsymbol j | \boldsymbol q_m
\rangle \langle \boldsymbol q_m | \boldsymbol k \rangle.
\label{limprob_ev}
\eea
This expression simplifies the numerical evolution considerably.
\subsection{$M\times M$ networks} \label{limprobsquare}
\begin{figure}[!ht]
\centerline{\includegraphics[clip=,width=0.9\columnwidth]{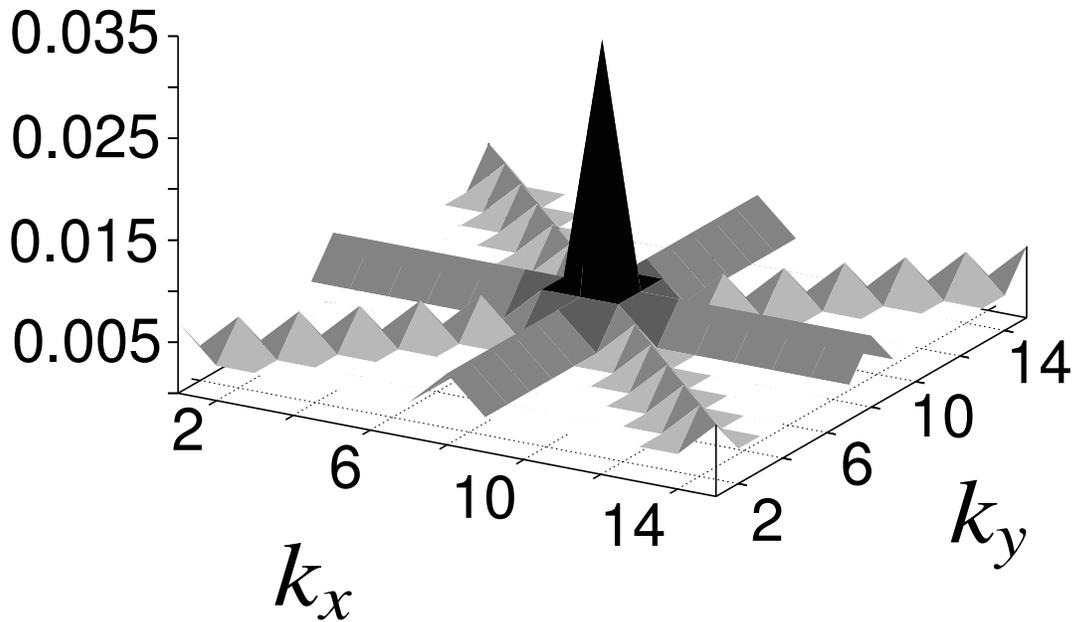}}
\caption{Limiting probabilities for a $15\times15$ cylinder, the walk starts at a central node.
}
\label{chicenter15x15}
\end{figure}

Let us first consider the behaviour of $\chi_{\boldsymbol{k,j}}$ taking as initial node $\boldsymbol{j}$ the centre of a $M\times M$ network, where $M$ is odd. The classical behaviour is diffusive and it leads to an equipartition among all nodes of the network, hence classically $\lim_{t\to\infty}\bar{p}(t)=1/{\mathcal{N}}$. Quantum-mechanically however, the situation is different. Figure \ref{chicenter15x15} shows the pattern of $\chi_{\boldsymbol{k,j}}$ for a $15\times15$ cylinder. There is a very pronounced global maximum at the centre. Other local maxima are to be found on the nodes placed on the diagonals of the structure and on the shortest paths which connect the centre to the four sides. The LPs of the other nodes produce a plateau which is approximately half of the local maximum. Furthermore, we find  the same result, up to our numerical precision, when we consider a square or a torus with the same initial condition.

\begin{figure}[!ht]
\centerline{\includegraphics[clip=,angle=90,width=0.9\columnwidth]{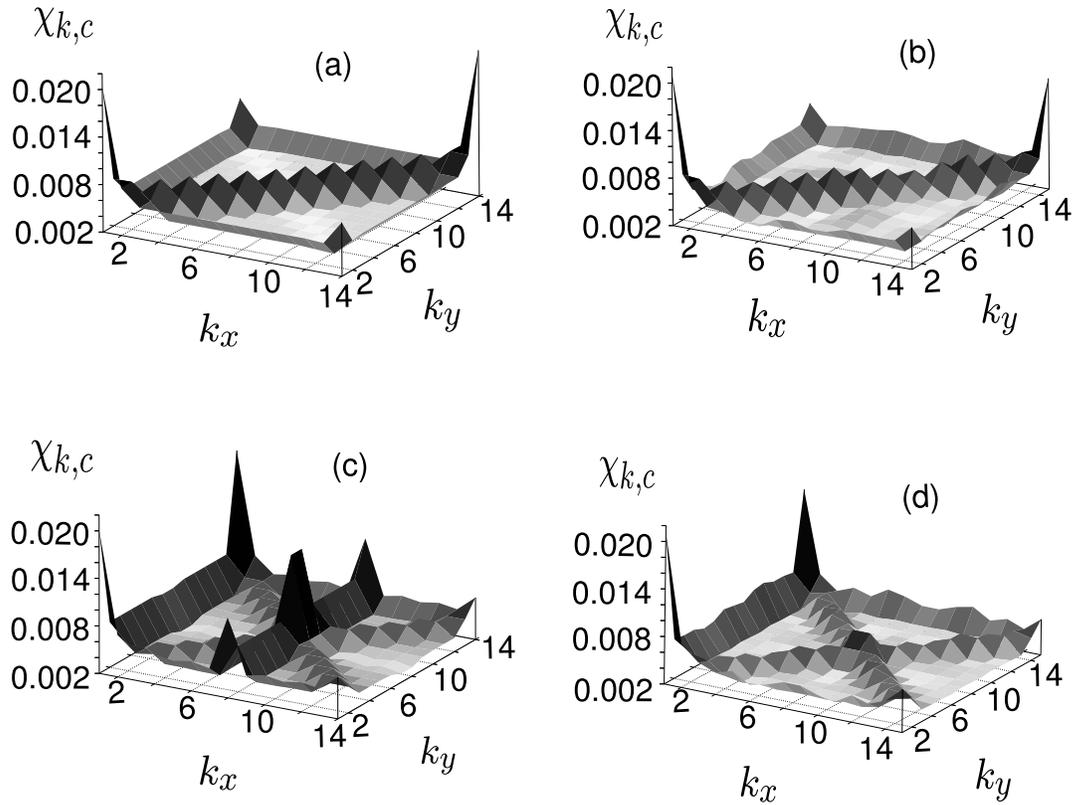}}
\caption{LP distributions for walks starting at $\boldsymbol{j}\equiv (1,1)$ for (a) the $14\times14$ square, (b) the $15\times15$ square, (c) the $14\times14$ cylinder and (d) the $15\times15$ cylinder.
}
\label{limtingprob}
\end{figure}

\begin{figure}[!ht]
\centerline{\includegraphics[clip=,angle=0,width=0.7\columnwidth]{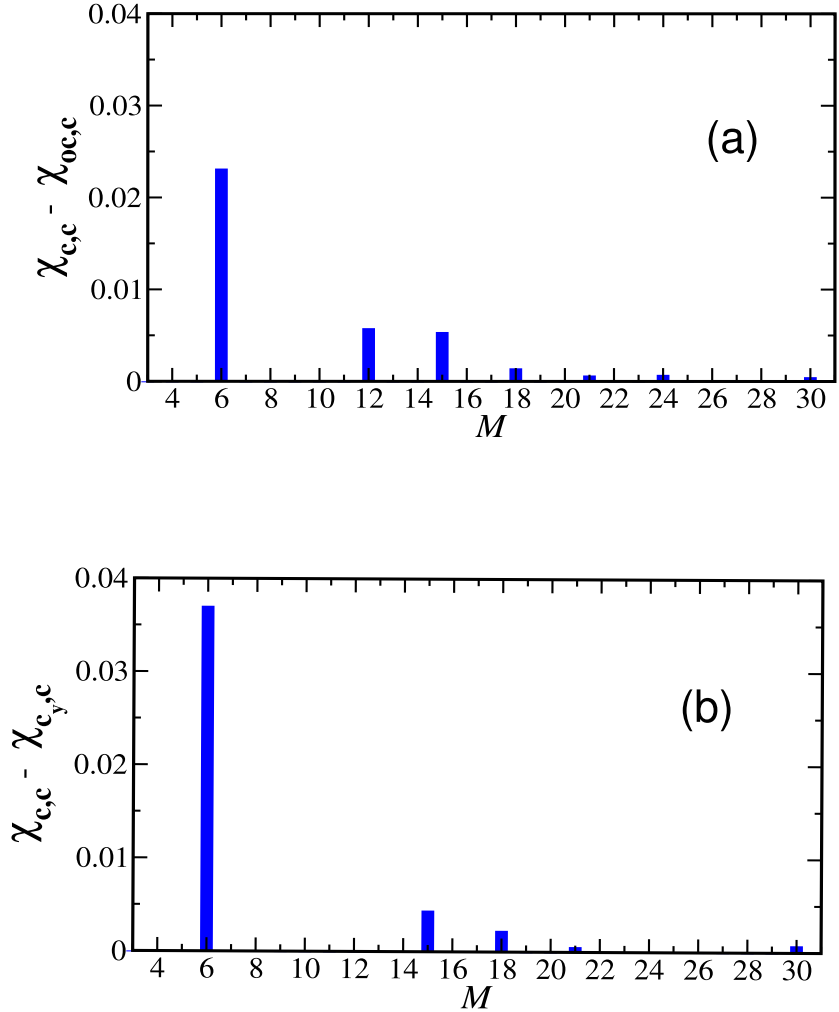}}
\caption{ (a) Differences between the LP for CTQWs that start at $\boldsymbol{c}=(1,1)$ to be at $\boldsymbol{c}$, $\chi_{\boldsymbol{c,c}}$, and to be at $\boldsymbol{oc}=(M,M)$, $\chi_{\boldsymbol{oc,c}}$ for OBC. (b) Differences between the LPs for CTQWs on a cylinder that start at $\boldsymbol{c}=(1,1)$ to be at $\boldsymbol{c}$, and to be at $\boldsymbol{c_y}=(1,M)$.  
}
\label{chicocsquare}
\end{figure}

~\\Now we place the initial excitation at one corner of the structure, say at $\boldsymbol{j}=(1,1)$. For the torus all nodes are topologically equivalent, so that the result is identical to the one displayed in Fig.~\ref{chicenter15x15}. We expect hence differences only in the case of squares and of cylinders.\\
Now, for square networks one intuitively expects the $\chi_{\boldsymbol{k,j}}$ to be symmetric with respect to the centre of the network, i.e., one expects the node $(j_x,j_y)$ and its mirror node, defined by $(M+1-j_x,M+1-j_y)$ to have the same LP. In general, this turns out to be true, but, there are some exceptions \cite{mvb2005a}, namely for $M=6, 12, 15, 18, 21, 24, 30, 36, \ldots$. In Fig.~\ref{limtingprob} we exemplify this behaviour. Thus the $14\times14$ square has a symmetric LP, shown in Fig.~\ref{limtingprob}(a), whereas the $15\times15$ square has an asymmetric LP, see Fig.~\ref{limtingprob}(b); note the differences along the borders and also the differences at the points $(1,1)$ and $(M,M)$ in Fig.~\ref{limtingprob}(b), compared to Fig.~\ref{limtingprob}(a). These differences will become clearer in the discussion of Fig.~\ref{chicocsquare}.       

~\\With cylindrical boundary conditions we find a similar situation. However, the symmetry is of another kind. We choose OBCs in the $y$-direction and PBCs in the $x$-direction. In Fig.~\ref{limtingprob}(c) we plot the LPs for the $14\times14$ cylinder. In this case there is a mirror symmetry on the rows with constant $k_x$, i.e., $\chi_{\boldsymbol{k,c}}$ is the same for the nodes $(k_x,k_y)$ and $(k_x,M-k_y)$. This is not true for the $15\times15$ cylinder, for which the LPs are given in Fig.~\ref{limtingprob}(d).\\Because of the PBC along the $x$-direction one expects the same LPs, for instance, at the second and the last nodes along the $x$-direction. This is indeed borne out by the figures. To render clearer these asymmetric behaviours, we plot in Fig.~\ref{chicocsquare}(a) as a function of $M$ the differences $\chi_{\boldsymbol{c,c}}-\chi_{\boldsymbol{oc,c}}$ for $\boldsymbol{c}=(1,1)$ and $\boldsymbol{oc}=(M,M)$. We obtain values different from 0 for $M=6,12,15,18,21,24,30,\ldots$. The situation for the cylinders is shown in Fig.~\ref{chicocsquare}(b), where we plot $\chi_{\boldsymbol{c,c}}-\chi_{\boldsymbol{c_y,c}}$ with $\boldsymbol{c_y}=(1,M)$. The asymmetries appear then for $M=6,15,18,21,30,\ldots$. Remarkably, in both cases, non-zero $\chi_{\boldsymbol{c,c}}-\chi_{\boldsymbol{c_y,c}}$ values are found only for $N$ values which are multiples of three. 

\begin{figure}[!ht]
\centerline{\includegraphics[clip=,angle=0,width=0.7\columnwidth]{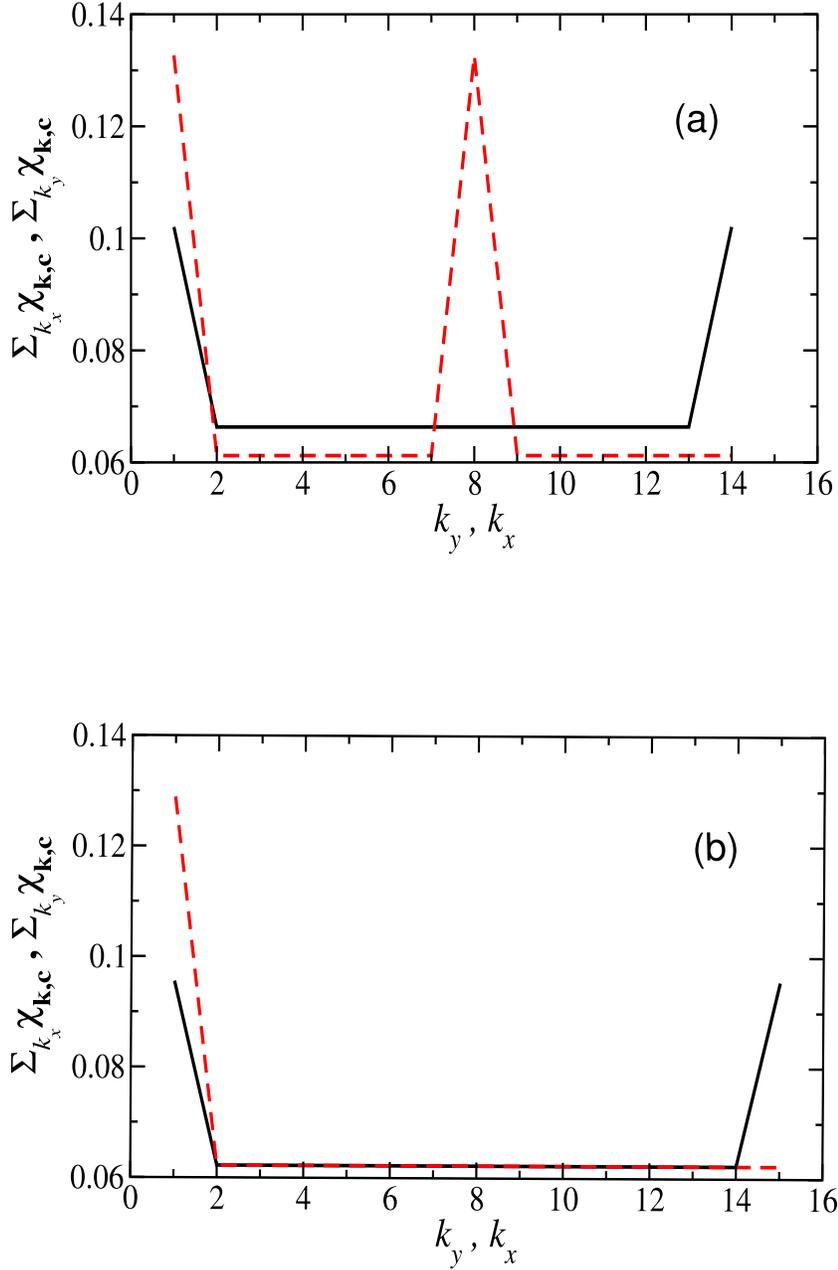}}
\caption{Comparison between the marginal LPs  to $\sum_{k_x}\chi_{\boldsymbol{k,c}}$ (full line) and $\sum_{k_y}\chi_{\boldsymbol{k,c}}$ (dashed line) on a $14\times14$ network (a) and on a $15\times15$ network (b), see text for details.  
}
\label{rowscolumns}
\end{figure}
~\\One can now sum the LPs along one of the two directions ($x$ or $y$) of the network. For a cylinder this leads after summing along the open $y$-direction to a ring-like behaviour and after summing along the $x$-direction with PBCs to a chain-like behaviour.
Figure \ref{rowscolumns}(a) shows the LP sums $\sum_{k_x=1}^M\chi_{\boldsymbol{k,c}}$ (full line) and $\sum_{k_y=1}^M\chi_{\boldsymbol{k,c}}$ (dashed line) for the $14\times14$ network with $\boldsymbol{k}=(k_x,k_y)$ and $\boldsymbol{c}=(1,1)$. Both curves have two maxima. For the chain the maxima occur at the extremes ($k_y=1$ and $k_y=14$), whereas for the ring they arise at $k_x=1$ and $k_x=8$.\\   
In Fig.~\ref{rowscolumns}(b) we plot the corresponding results for the $15\times15$ network. For the chain one finds two equal maxima at the extremal points and a low, equal probability in the middle. For the ring, on the other hand, we find just one maximum  at $k_x=1$ whose value is $(2M-1/M^2)\approx 0.1288$ and a plateau $(M-1)/M^2\approx 0.06222$ for $k_x\neq1$. 
Here, distinct from Fig.~\ref{rowscolumns}(a), the plateau has the same value in both cases (chain or ring).

\subsection{$M\times N$ networks} \label{limprobret}
\begin{figure}[!ht]
\centerline{\includegraphics[clip=,angle=0,width=0.9\columnwidth]{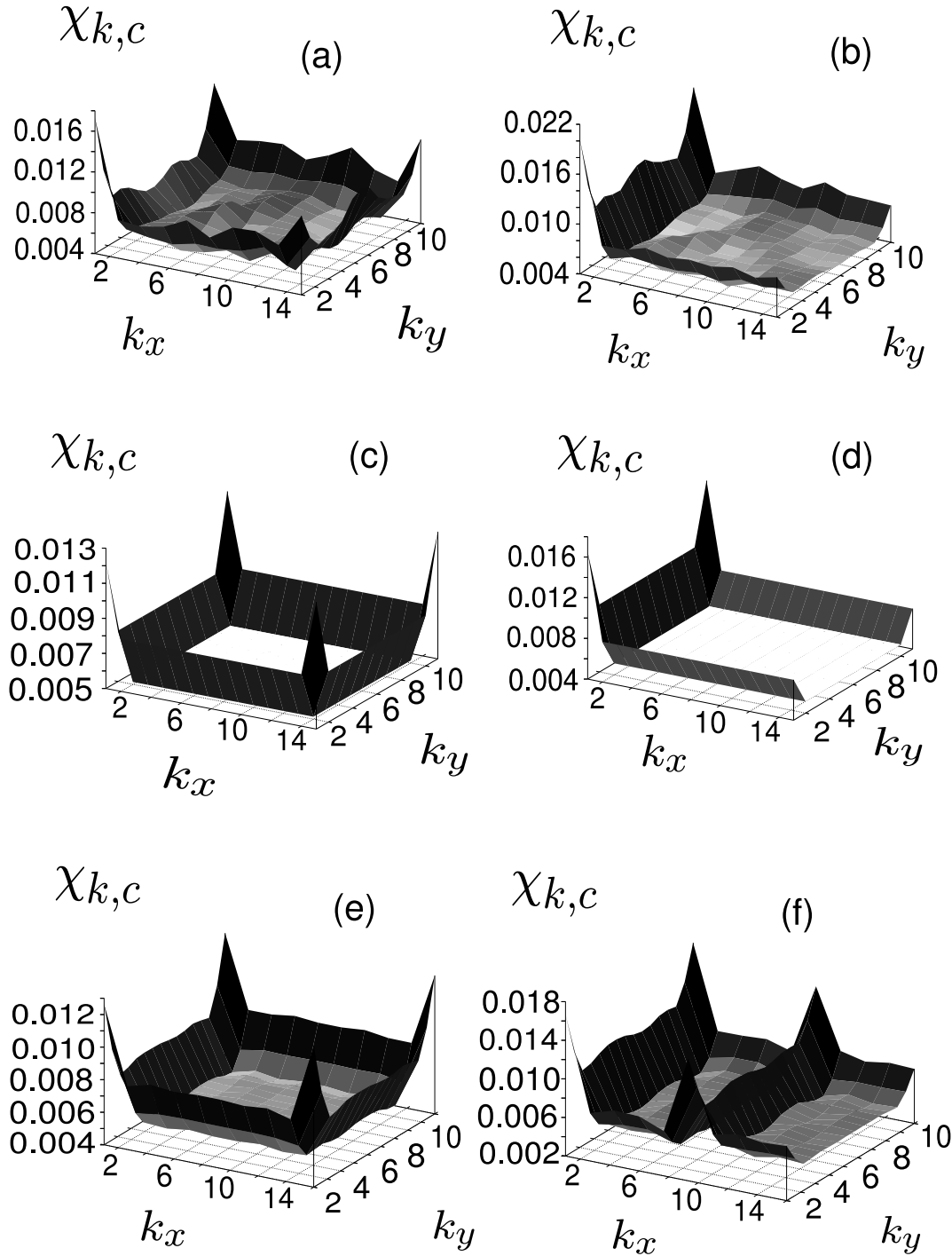}}
\caption{LPs for (a) the $15\times10$ rectangle, (b) the $15\times10$ cylinder, (c) the $15\times 11$ rectangle, (d) the $15 \times 11$ cylinder,(e) the $16\times10$ rectangle, (f) the $16\times10$ cylinder.  
}
\label{limprobrect}
\end{figure}

\begin{figure}[!ht]
\centerline{\includegraphics[clip=,angle=0,width=0.675\columnwidth]{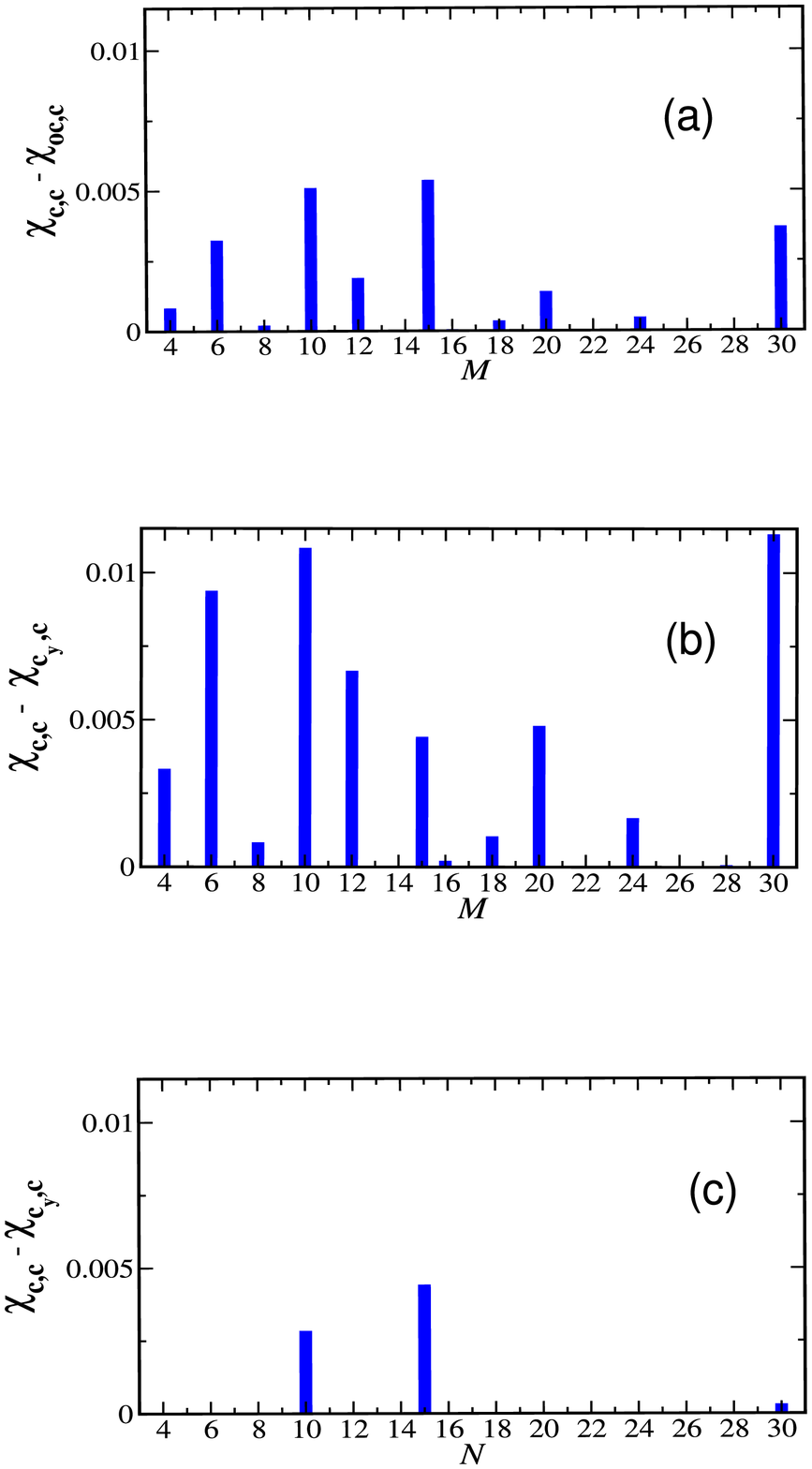}}
\caption{(a) Rectangles $M\times N$ with $N=15$ and varying $M$: Differences between the LPs for CTQWs that start at $\boldsymbol{c}=(1,1)$ to be at $\boldsymbol{c}$, $\chi_{\boldsymbol{c,c}}$, and to be at $\boldsymbol{oc}=(M,N)$, $\chi_{\boldsymbol{oc,c}}$. (b) Cylinders $M\times N$ with $N=15$  and varying $M$: Differences between the LPs for CTQWs that start at $\boldsymbol{c}=(1,1)$ to be at $\boldsymbol{c}$, and to be at the node $\boldsymbol{c_y}=(1,N)$. (c) Cylinders $M\times N$ with $M=15$  and varying $N$: Differences between the LPs for CTQWs that start at $\boldsymbol{c}=(1,1)$ to be at $\boldsymbol{c}$, and to be at the node $\boldsymbol{c_y}=(1,N)$.
}
\label{chicoc}
\end{figure}

We extend now the analysis of section \ref{limprobsquare} to general $M\times N$ networks with $M\neq N$, considering again different boundary conditions. 
Figure \ref{limprobrect} shows examples of symmetric and asymmetric patterns for networks of sizes $15\times10$ [(a) and (b)], $15\times 11$ [(c) and (d)]  and $16\times 10$ [(e) and (f)], the networks being either rectangles [(a), (c) and (e)] or cylinders [(b), (d) and (f)]. For rectangles, OBCs, the global maximum is at the initial site and there are pronounced maxima of the LPs along the edges of the structure; this is similar to the situation for $M\times M$-networks shown in Fig.~\ref{limtingprob}. However, here we do not find local maxima along the diagonals. Clearly, the LPs for the $15\times 10$ rectangle show more structure [Fig.~\ref{limprobrect}(a)] than the other two networks. Thus, the $15\times 11$ network [Fig.~\ref{limprobrect}(c)] is symmetric and its LPs take only three different values; the $16\times10$ network [Fig.~\ref{limprobrect}(e)] is very regular and its LPs are concentrated on the peripheral nodes.\\Figures \ref{limprobrect}(b), (d) and (f) present the corresponding cylindrical networks. The LPs for the $15\times10$ network have their global maximum at $\boldsymbol{c}\equiv(1,1)$ and, again, there is no mirror symmetry for the nodes $(k_x,k_y)$ and $(k_x,N-k_y)$. The $\chi_{\boldsymbol{k,c}}$ for the $15\times11$ cylinder [Fig.~\ref{limprobrect}(d)] are very regular. One observes two equally large peaks at $\boldsymbol{c}\equiv(1,1)$ and at $\boldsymbol{c_y}\equiv(1,11)$. For the other points of the row $k_x=1$, the LPs are equal. The same holds for the nodes of the cylinder (at $k_y=1$ and at $k_y=11$) and for the other remaining nodes of the network. Figure \ref{limprobrect}(f) displays $\chi_{\boldsymbol{k,c}}$ for the $16\times10$ cylinder; the LP distribution is quite different from the previous two patterns. As in Fig.~\ref{limtingprob}(c) there appear two "crests", at $k_x=1$ and at $k_x=8$, a result due to the PBCs along the $x$-axis and to the fact that here $M$ is even ($M=16$).\\In order to systematically analyse the asymmetries of $\chi_{\boldsymbol{k,j}}$ in rectangular networks, we now fix $N$ to be $N=15$ and vary $M$, taking $4\leq M\leq 30$. Figure \ref{chicoc}(a) shows for rectangles the difference between the LPs on the initial corner $\chi_{\boldsymbol{c,c}}$ and on the corner $\chi_{\boldsymbol{oc,c}}$. In the analysed range, $4\leq M\leq 30$, the value $\chi_{\boldsymbol{c,c}}-\chi_{\boldsymbol{oc,c}}$ displays varying patterns. For $4 \leq M\leq13$ one can associate $\chi_{\boldsymbol{c,c}}-\chi_{\boldsymbol{oc,c}}=0$ to odd values of $M$ and $\chi_{\boldsymbol{c,c}}-\chi_{\boldsymbol{oc,c}}\neq0$ to even values of $M$; for larger $M$ the situation becomes more complex. Figure \ref{chicoc}(b) displays the situation for cylinders with $N=15$ and $4\leq M\leq 30$. Here we plot $\chi_{\boldsymbol{c,c}}-\chi_{\boldsymbol{c_y,c}}$, with $c_y\equiv (1,N)$. Figure \ref{chicoc}(c) shows the situation for cylinders with $M=15$ and $4\leq N\leq 30$. This last case looks quite regular, with non-vanishing values only for $N=10, 15,$ and $30$. Hence, Figs.~\ref{chicoc}(b) and \ref{chicoc}(c) ($N<M$) show for cylinders that changes of the radius lead to more asymmetric situations than changes of the length.

~\\We close by briefly reviewing some of the details of the calculations of
the LPs, which give an indication of the origin of the asymmetries. As we
have seen in Sec.~\ref{bound}, the Hamiltonian of the problem separates in the two
directions. Using this fact, it is easy to show that the transition
probabilities $\pi_{{\boldsymbol k},{\boldsymbol j}}(t)$ can be written as the product of
the two separate probabilities for each direction, i.e.
\begin{equation}
\pi_{{\boldsymbol k},{\boldsymbol j}}(t) = \pi_{k_x, j_x}(t) \ \pi_{k_y, j_y}(t)
= |\alpha_{k_x, j_x}(t)|^2 \ |\alpha_{k_y, j_y}(t)|^2,
\end{equation}
with $\alpha_{k_x, j_x}(t) = \sum_{\theta_x} \exp(-i\lambda_{\theta_x}t)
\langle k_x | \Psi_{\theta_x} \rangle \langle \Psi_{\theta_x} | j_x
\rangle$
, and similarly for the $y$-direction.\\
Now, according to Eq.~(\ref{limprob_ev}), the LPs are given by
\begin{eqnarray}
\chi_{{\boldsymbol k},{\boldsymbol j}} &=& \lim_{T\to\infty}\frac{1}{T} \int_0^T dt \
\pi_{k_x,
j_x}(t) \ \pi_{k_y, j_y}(t) \nonumber \\
&=& \sum_{\theta_x,\theta'_x,\theta_y, \theta'_y} F_{\boldsymbol{k,j}}
\lim_{T\to\infty} \frac{1}{T} \int_0^T dt \
\exp\left[-it(\lambda_{\theta_x}
- \lambda_{\theta'_x} + \lambda_{\theta_y} -
  \lambda_{\theta'_y})\right],
\label{xi_asymm}
\end{eqnarray}
where $F_{\boldsymbol{k,j}}$ is a time independent function, which depends
on the eigenstates associated with $\theta_x$, $\theta'_x$, $\theta_y$,
and $\theta'_y$. Because of the limit in the time integral in Eq.~(\ref{xi_asymm}) there
 are only contributions to $\chi_{{\boldsymbol k},{\boldsymbol j}}$ if a value
$(\lambda_{\theta_x}
- \lambda_{\theta'_x})$ for the $x$-direction has a counterpart
  $-(\lambda_{\theta_y}
- \lambda_{\theta'_y})$ in the $y$-direction.\\A careful analysis of the differences $(\lambda_{\theta_x}
- \lambda_{\theta'_x})$ indicates where the asymmetries stem from. For
finite
 chains we obtain $(\lambda_{\theta_x} - \lambda_{\theta'_x}) = 2\cos
\theta'_x - 2\cos
\theta_x$. For simplicity we consider now finite $M\times M$ networks
with OBCs, see \cite{mvb2005a}, because then the eigenvalues are the same in both
directions. It turns out that for 
$\theta_x\neq\theta'_x$ the value $(\lambda_{\theta_x} - \lambda_{\theta'_x})$ appears only once or twice for all
symmetric cases. However, for the asymmetric cases, some of the $(\lambda_{\theta_x} -
\lambda_{\theta'_x})$ values (again for $\theta_x\neq\theta'_x$) appear more than twice. Therefore,
there are more contributions to $\chi_{{\boldsymbol k},{\boldsymbol j}}$ in the
asymmetric cases than in the symmetric cases. Given that we have all eigenvalues analytically, we can in principle study the asymmetries fully analytically. However, a complete
analysis, including different boundary conditions, requires a thorough
investigation of all possible differences of eigenvalues. This is a quite
extensive task and is clearly beyond the scope of
this paper. Thus, a detailed analysis of the asymmetries will be given
elsewhere. 
\section{Conclusion} \label{conclusion}
To summarise, we have systematically analysed the role of different boundary conditions on the quantum mechanical transport on two dimensional graphs. We have kept the systems as simple as necessary to still highlight the complex behaviour of the quantum mechanical transport. In the long time average of the transition probability distribution $\chi_{\boldsymbol{k,j}}$ we found asymmetries, depending on the size {\sl and} the BCs of the graph. Although graphs with toroidal topologies have a perfectly symmetric $\chi_{\boldsymbol{k,j}}$, this does not have to be the case for graphs of the same size with open BCs. For networks with an equal number of nodes in both directions as well as
for networks with an unequal number, the asymmetries are best visualised
by comparing the LP of the initial node to the one of an appropriate
"mirror node". Here we found significant differences for the specific
sizes of the network depending on the BCs. We further showed that summing along one side of the graph leads to the known behaviours for quantum mechanical transport over chains (OBCs) or rings (PBCs).       
\section*{Acknowledgments}

This work was supported by a grant from the Ministry of Science, Research
and the Arts of Baden-W\"urttemberg (AZ: 24-7532.23-11-11/1). Further
support from the Deutsche Forschungsgemeinschaft (DFG) and the Fonds der
Chemischen Industrie is gratefully acknowledged.

\section*{References}

\bibliographystyle{iopart-num}

\providecommand{\newblock}{}

\end{document}